\documentclass[11pt,letterpaper]{article}
\usepackage[margin=1in]{geometry}
\usepackage[T1]{fontenc}
\usepackage{lmodern}
\usepackage{microtype}
\usepackage{amsmath,amssymb,bm}
\usepackage{graphicx}
\usepackage{cite}
\usepackage[hidelinks,hyperfootnotes=false]{hyperref}
\usepackage[capitalize]{cleveref}

\crefname{figure}{Fig.}{Figs.}
\crefname{equation}{Eq.}{Eqs.}
\newcommand{\mathd}{\mathrm{d}}

\title{Dissipation fingerprints of chiral topological superconductors}
\author{%
  Hailong Li\(^{1,*,\dagger}\), Qing Yan\(^{1,\dagger}\),
  Hua Jiang\(^{1}\), and X. C. Xie\(^{1,2,3,*}\)\\[0.8em]
  \parbox{0.95\textwidth}{\centering\small
    \(^{1}\)Interdisciplinary Center for Theoretical Physics and Information Sciences
    (ICTPIS), Fudan University, Shanghai, 200433, China\endgraf
    \(^{2}\)International Center for Quantum Materials, School of Physics,\\
    Peking University, Beijing, 100871, China\endgraf
    \(^{3}\)Hefei National Laboratory, Hefei, 230088, China\endgraf
  }%
}
\date{}
\hypersetup{%
  pdftitle={Dissipation fingerprints of chiral topological superconductors},
  pdfauthor={Hailong Li, Qing Yan, Hua Jiang, X. C. Xie},
  pdfkeywords={Quantum transport, Energy dissipation, Topological superconductors, Chiral Majorana edge modes}
}

\begin{document}
\maketitle
\begingroup
\renewcommand{\thefootnote}{\fnsymbol{footnote}}
\footnotetext[1]{Corresponding authors (Hailong Li, email: \href{mailto:hailongli@fudan.edu.cn}{hailongli@fudan.edu.cn}; X. C. Xie, email: \href{mailto:xcxie@pku.edu.cn}{xcxie@pku.edu.cn})}
\footnotetext[2]{These authors equally contributed to this work.}
\endgroup

\begin{abstract}
Chiral topological superconductors (TSCs) in two dimensions are predicted to host chiral Majorana edge modes, yet their unambiguous experimental identification remains a major challenge. 
The half-quantized electrical conductance plateau was long considered {a potential signature}, but recent findings that trivial phases can mimic this behavior have cast doubt on its validity. 
Here, we show that spatially resolved energy dissipation serves as a robust fingerprint to resolve this ambiguity. 
Contrary to conventional superconductors where dissipation arises from vortex motion, we demonstrate that even static vortices in chiral TSCs induce measurable edge-state dissipation via inelastic scattering with vortex core states. 
By analyzing hybrid quantum anomalous Hall--TSC junctions, we find that while phases with Chern numbers $N=0, 1, 2$ all exhibit identical half-quantized electrical conductance, they present {distinct spatial dissipation profiles that can help identify the chiral Majorana modes}.
As a complementary consistency check, we show that the intrinsic transverse thermal Hall response remains quantized in an ideal chiral-edge geometry.
Together, our results establish dissipation imaging as {a conceptually distinct approach to probing} chiral topological superconductivity.
\end{abstract}

\begin{center}\footnotesize\textbf{Keywords.} Quantum transport, Energy dissipation, Topological superconductors, Chiral Majorana edge modes
\end{center}

\section{Introduction}

The interplay between superconductivity and topology gives rise to a novel class of quantum phases known as topological superconductors (TSCs)~\cite{kitaevUnpairedMajoranaFermions2001,hasanColloquiumTopologicalInsulators2010,qiTopologicalInsulatorsSuperconductors2011,aliceaNewDirectionsPursuit2012,nadj-perge2014ScienceMajoranaFermionsFeChain,mourikSignaturesMajoranaFermions2012,sunJiaJF2016PRLMajoranaZeroMode,zhangDingHong2018ScienceIronFeTopologicalSuperconductivity,wangDingH2018ScienceEvidenceMajoranaIronTopologicalSuperconductivity}.
Two-dimensional (2D) chiral topological superconductors represent a distinctive subclass breaking time-reversal symmetry, where the nontrivial bulk topology gives rise to gapless chiral Majorana edge modes~\cite{FuliangKane2009PRL,AkhmerovBeenakker2009PRL,Beenakker2013annurev-conmatphys,Elliott2015RevModPhys,qiChiralTopologicalSuperconductor2010,yanqing2021CommunPhysAxionMajorana}.
Due to their connection with non-Abelian statistics~\cite{lianTopologicalQuantumComputation2018a,zhouNonAbelianOperationChiral2019,linNonAbelianOperationScattering2022,wuRecentProgressNonAbelian2023,huangUnifiedModelNonAbelian2026} and potential applications in fault-tolerant quantum computation, the realization of chiral TSCs has become one of the central goals and major challenges in condensed matter physics~\cite{Nayak2008RevModPhys,Campbell2017naturereviewQC,lianTopologicalQuantumComputation2018a}.
In particular, half-quantized electrical conductance plateaus in hybrid quantum anomalous Hall (QAH)--TSC junctions have long been viewed as the hallmark of chiral Majorana edge modes~\cite{chungConductanceNoiseSignatures2011a,wangChiralTopologicalSuperconductor2015}, yet recent theory and experiment have demonstrated that similar plateaus can arise from mechanisms unrelated to an $N=1$ chiral TSC or an isolated chiral Majorana mode~\cite{Chenchuizhen2017PRBdomainwall,JiWenjieWenxiaogang2018PRL,HuangYingyi2018PRB,kayyalhaAbsenceEvidenceChiral2020,udayNonMajoranaOriginHalfinteger2025,royHalfintegerThermalConductance2026}, casting doubt on this criterion.
Consequently, {a conceptually distinct experimental} fingerprint of chiral TSCs remains elusive.

Topological edge states are inherently robust against local perturbations and are commonly believed to support energy-dissipationless transport~\cite{changColloquiumQuantumAnomalous2023}.
In superconductors, dissipationless transport is reinforced by the macroscopic phase coherence of the condensate. 
It is thus natural to expect that TSCs, which unify these two paradigms, would likewise host ideal dissipationless transport~\cite{WangjingZhangSC2017NatMaterTopoCondensedMatter}. 
However, this idealized picture has been challenged by recent imaging experiments in non-superconducting topological phases such as quantum Hall (QH) and QAH insulators, where chiral edge states exhibit measurable energy dissipation~\cite{slizovskiyCoolingChiralHeat2017,margueriteImagingWorkDissipation2019,zhangDissipationResistanceImaging2020,fangThermalDissipationQuantum2021,liEmergentEnergyDissipation2024,yanRulesDissipationlessTopotronics2024,yanDissipationlessTopotronics2025,liNewAdvancesDissipationless2025}.
{Recent studies have linked local heat dissipation in QAH systems to effective non-Hermitian transport arising from coexisting chiral and diffusive edge modes~\cite{zhouNonHermitianTopology2026}.
Near-field radiative energy loss generated by local shot noise has also been identified in topological insulators~\cite{weiHiddenDissipation2026}.} It has also been show that dissipation can induce first order phase transition in normal surperconductors~\cite{liuDissipationDrivenFirstOrder2026}.
Despite the intense interest in chiral TSCs, however, the question of whether and how dissipation manifests in these systems has remained largely unexplored.

\begin{figure}[!htbp]
  \centering
  \includegraphics[width=0.88\linewidth]{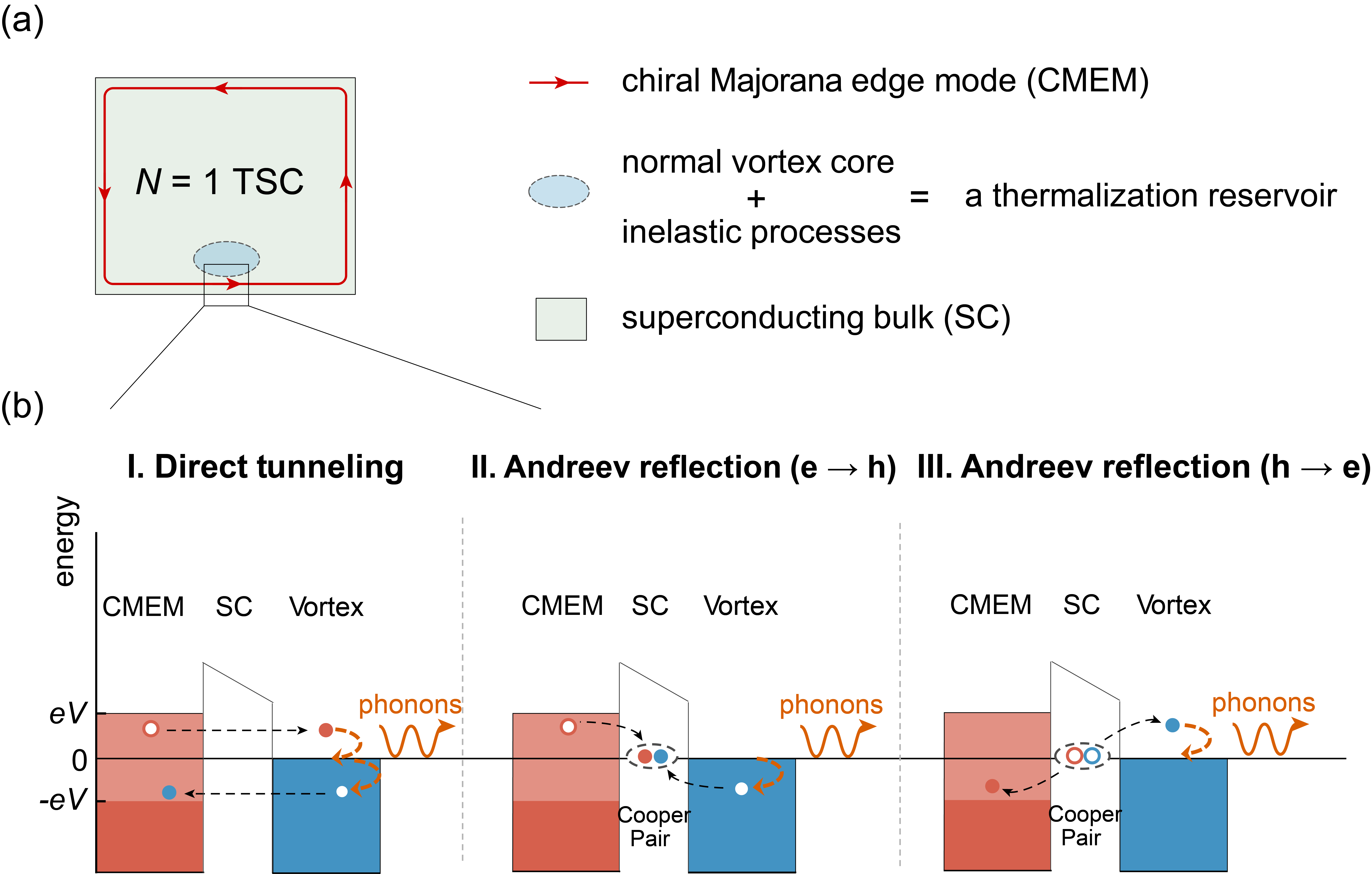}
  \caption{
(a) A two-dimensional chiral topological superconductor (TSC) with Chern number $N=1$ hosts a single gapless chiral Majorana fermion along its edge. The blue shaded region marks a normal vortex core where metallic states reside.
(b) Schematic illustration of the scattering processes between chiral edge modes and a normal vortex core under a small voltage bias $V$ with $eV\!\ll\!\Delta$ (the superconducting gap). Single-particle tunneling, electron-type Andreev reflection, and hole-type Andreev reflection induce a nonequilibrium state, which subsequently relaxes via inelastic scattering to phonons, thereby generating local energy dissipation. 
}
  \label{Fig1}
\end{figure}

Vortices play a pivotal role in superconductors. In conventional superconductors, energy dissipation arises from the motion of vortices: an electric field induced by a moving vortex drives normal currents through its core, producing resistance and dissipation~\cite{bardeenTheoryMotionVortices1965}. By contrast, when vortices are pinned, supercurrents flow around them without resistance, resulting in dissipationless transport. This standard picture implicitly assumes that static vortices are irrelevant for energy dissipation.

Here, we challenge this conventional view by demonstrating that dissipation is not only present in chiral TSCs but serves as a crucial identifier of their topological nature. 
In contrast to conventional superconductors, we identify a mechanism where even static (pinned) vortices induce measurable dissipation via inelastic scattering between chiral edge modes and vortex core states.
Despite undermining the expectation of dissipationless Majorana transport, we show that this dissipation offers an opportunity to resolve the long-standing ambiguity of half-quantized electrical conductance. By analyzing realistic QAH--TSC junctions, we find that while topological phases with Chern numbers $N\!=\!0, 1, 2$ exhibit identical half-quantized electrical conductance, they display unique spatial dissipation patterns.
As a complementary consistency check, we further show that the intrinsic thermal Hall conductance remains quantized, consistent with the underlying bulk topology. Our work thus {presents an approach} that turns the unexpected dissipation into {a fingerprint for probing} chiral topological superconductivity.

\begin{figure}[!htbp]
  \centering
  \includegraphics[width=0.96\linewidth]{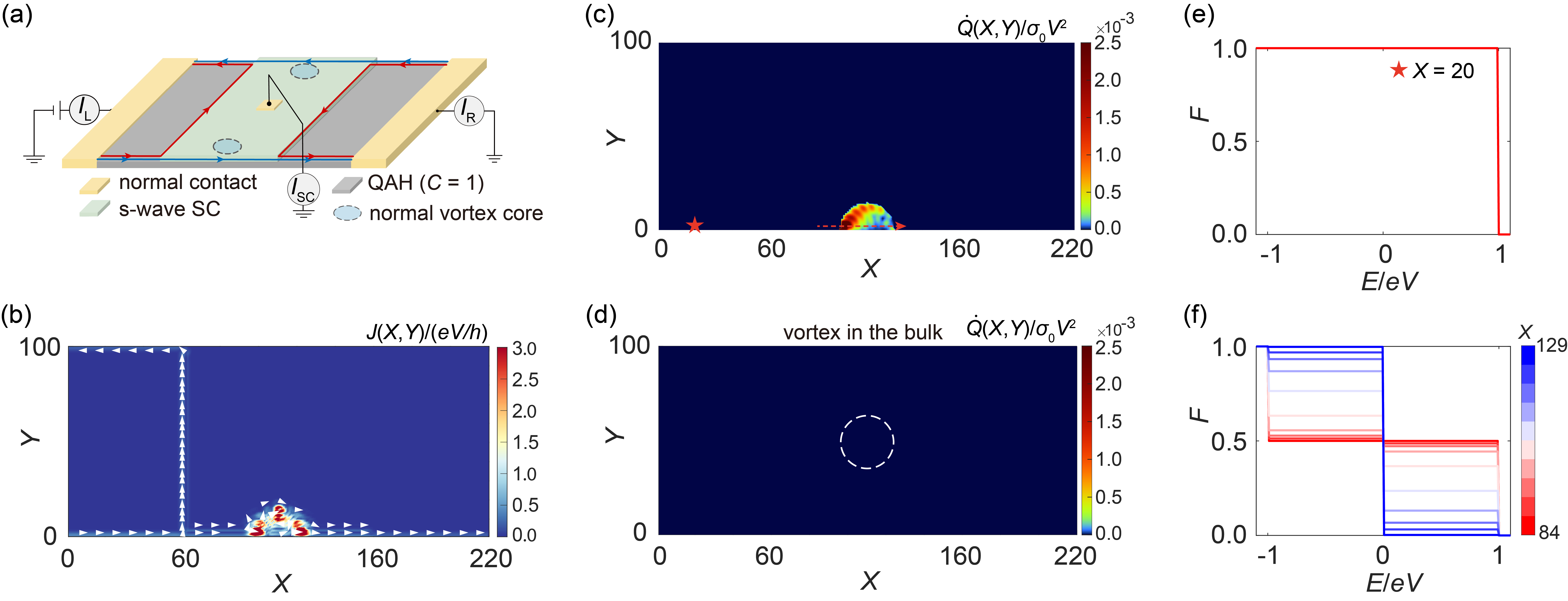}
  \caption{
(a) Schematic of a quantum anomalous Hall (QAH) insulator proximitized by an $s$-wave superconductor, forming a chiral topological superconductor (TSC) with Chern number $N\!=\!1$ in the central region. {Two static vortices are placed near the bottom
and top edges of the TSC, respectively.} A small bias voltage $V$ is applied between the source and drain, while the TSC is grounded. Red and blue arrows indicate the configuration of chiral edge states.
(b) Quasiparticle flow in the device under bias $V$. Only the lower-edge vortex lies on the trajectory of the net quasiparticle flow.
(c) Spatial distribution of the local energy dissipation power $\dot{Q}(X,Y)$ in units of $\sigma_0 V^2$ for a two-vortex configuration, with one vortex near the bottom edge and the other near the top edge. Dissipation occurs only at the bottom-edge vortex, while no visible signal appears at the top-edge vortex.
(d) Control calculation for a single vortex positioned at the center of the TSC region.
(e)(f) Local occupation function $F$ at the red star and along the dashed red line marked in panel (c), illustrating the nonequilibrium behavior near the dissipation hotspot.
  }
  \label{Fig2}
\end{figure}

\section{Physical picture of dissipation mechanism}
Two-dimensional chiral TSCs, analogous to QH and QAH insulators, are characterized by an integer topological invariant—the Chern number $N$~\cite{Altland-ZirnbauerSymmetryClassificationPRB}. A chiral TSC with Chern number $N$ hosts $N$ chiral Majorana edge modes (CMEMs), each carrying half the degrees of freedom of a normal electronic mode. As illustrated in~\cref{Fig1}(a), a single gapless CMEM appears at the boundary when $N=1$.
When a small source--drain bias is applied, edge currents are carried by CMEMs. We then consider {a static Abrikosov-type vortex, or more generally a superconductivity-suppressed core region,} pinned near the boundary (blue-shaded region in~\cref{Fig1}(a)), whose core hosts localized normal states.
When a CMEM passes this core, scattering processes---such as single-particle tunneling and Andreev reflection---can occur between the edge quasiparticles and the core states (\cref{Fig1}(b)).
Notably, although these scattering processes are elastic, they generate a nonequilibrium distribution of quasiparticles in the vortex core, consisting of high-energy electrons and holes. 
Through inelastic scattering—primarily mediated by electron-phonon interactions—these quasiparticles relax via phonon emission, thereby producing local energy dissipation within the vortex.
Recent advances in nanoscale thermal imaging have made it feasible to directly probe energy dissipation~\cite{halbertalNanoscaleThermalImaging2016,zhouScanningSQUIDontipMicroscope2023}, offering a potential route to identify chiral topological superconducting phases.
To substantiate our proposed mechanism, we next numerically investigate the energy dissipation within the steady-state transport in a mesoscopic device hosting CMEMs and vortices.

\section{Spatial selectivity of vortex-induced dissipation}
Motivated by the theoretical proposal that chiral TSC phases can be realized by placing an $s$-wave superconductor in proximity to a QAH insulator~\cite{qiChiralTopologicalSuperconductor2010}, we consider a device setup illustrated in~\cref{Fig2}(a).
The system is described within the Bogoliubov-de Gennes (BdG) formalism, with the Hamiltonian defined as $\mathcal{H}_{\mathrm{BdG}}(\bm{p}) = \Psi_{\bm{p}}^{\dagger} H_{\mathrm{BdG}}(\bm{p}) \Psi_{\bm{p}} / 2$~\cite{qiChiralTopologicalSuperconductor2010, yanqing2019PRBCMEMs}, where
\begin{equation}
H_{\mathrm{BdG}}(\bm{p}) =
\begin{pmatrix}
h_{\mathrm{QAH}}(\bm{p}) - \mu_s & i \Delta \sigma_y \\
- i \Delta^{\ast} \sigma_y & - h_{\mathrm{QAH}}^{\ast}(-\bm{p}) + \mu_s
\end{pmatrix}\label{eq:BdG}
\end{equation}
and $\Psi_{\bm{p}} = \left(a_{\bm{p} \uparrow}, a_{\bm{p} \downarrow}, a^{\dagger}_{-\bm{p} \uparrow}, a^{\dagger}_{-\bm{p} \downarrow}\right)^{\mathrm{T}}$ is the basis.
Here, $h_{\mathrm{QAH}}(\bm{p}) = Ap_x \sigma_x + Ap_y \sigma_y + (m + B p^2) \sigma_z$ is the Qi-Wu-Zhang model for a QAH insulator~\cite{QiWuZhang2006PRBQAHmodel}.
The proximity-induced pairing potential $\Delta$ takes different values in different regions: it vanishes in the QAH and vortex regions, but remains finite in the superconducting region. Distinct superconducting phases with Chern number $N = 0$, $1$, or $2$ can be realized by tuning material parameters such as $m$, $B$, and $\mu_s$ (see Supplementary Note~1). In the superconducting region, the chemical potential $\mu_s$ is fixed by grounding.
Further details on the material parameters are provided in Supplementary Materials.

{We construct an $N\!=\!1$ TSC junction with two static vortices, one near the bottom edge and the other near the top edge, as illustrated in~\cref{Fig2}(a).}
In principle, one QAH edge state can be viewed as two CMEMs since a QAH phase with Chern number $C\!= \!1$ is topologically equivalent to an $N \!= \!2$ TSC~\cite{Altland-ZirnbauerSymmetryClassificationPRB,qiChiralTopologicalSuperconductor2010}.
Therefore, the chiral edge states in the QAH region split into two CMEMs at the TSC boundary (\cref{Fig2}(a)).
We then compute the steady-state edge transport of the junction within the BCS mean-field framework.
When a subgap voltage $eV\!\ll\!\Delta$ is applied between the source and drain, only the gapless CMEMs are involved.
In this framework, the quasiparticle current can be obtained from the continuity equation and numerically evaluated using the non-equilibrium Green's function (NEGF) method~(see Supplementary Note~3).
The spatial distribution of the quasiparticle current driven by net electron injection is shown in~\cref{Fig2}(b), where electrons are injected from the lower-left corner of the QAH region and subsequently split into two chiral Majorana modes at the domain wall.
At the bottom edge, the propagating quasiparticles encounter a vortex located in the TSC region.

{Phonon-mediated quasiparticle relaxation and vortex-assisted cooling have been studied microscopically in superconductors~\cite{savichQuasiparticleRelaxation2017,samokhvalovElectronPhonon2022}. Motivated by these results, we approximate a relaxing vortex core by a normal region coupled to a fictitious thermalization reservoir with self-energy $\Sigma_{\varphi}$~\cite{vaitkusButtikerProbesRecursive2018,maianiConductanceMatrixSymmetries2022}. Edge--core scattering injects nonequilibrium quasiparticles, while electron--phonon coupling transfers their excess energy to the lattice. This effective description assumes sufficiently dense or broadened core states and efficient equilibration, which captures the dissipative role of the core without resolving the full vortex texture.}
{Related B\"{u}ttiker-probe formulations have been used to describe dephasing in semiconductor--superconductor structures~\cite{mortensenDephasingSemiconductorSuperconductor2000} and to study Majorana bound states in the presence of inelastic scattering~\cite{mishraMajoranaFermionInduced2024}.}
At zero temperature, the total energy dissipation is entirely accounted for by the energy transferred to the thermalization bath.
The power of energy dissipation can be quantitatively evaluated using the NEGF formalism (see Supplementary Note~4), as
{
\begin{equation}
  \dot{Q}_p  =  \frac{1}{h} \sum_{\substack{
    \alpha \in \{\mathrm{e}, \mathrm{h}\}\\
    q}}\int \mathd \omega(\omega-\mu_p^{\mathrm{e}} ) T_{pq}^{\mathrm{e} \alpha} (\omega) [f_q^{\alpha} (\omega) -
  f_p^{\mathrm{e}} (\omega)]\label{eq:energy}
\end{equation}
}
where $p(q)$ labels all contacts including source, drain and the fictitious leads within the vortex.
{Here $\dot Q_p>0$ denotes heat absorbed by contact $p$, and the energy integral extends over the full real axis.} 
$\mu_p^{\mathrm{e}}\!=\!\mu_p-\mu_s$ is the chemical potential of electrons in contact $p$. 
Note that all energies $\omega$ are measured with respect to $\mu_s$.
$T_{pq}^{\mathrm{e} \alpha}$ is the transmission coefficient describing a particle of type $\alpha$ ($\!\in\!\{\mathrm{e},\mathrm{h}\}$) incident from contact $q$ to be transmitted to contact $p$ as an electron, which exactly accounts for the contributions from both the normal tunneling and Andreev scattering pictured in~\cref{Fig1}(b).
$f^{\alpha} (\omega)$ is the Fermi-Dirac distribution for particles of type $\alpha$, where $f^{\mathrm{e}/\mathrm{h}}_p (\omega)= \left[1+\exp\left({(\omega-\mu_p^{\mathrm{e}/\mathrm{h}})}/{k_{\mathrm{B}}T}\right)\right]^{-1}$ with $\mu_p^{\mathrm{e}} = -\mu_p^{\mathrm{h}}$.

{Following~\cref{eq:energy}, we observe energy dissipation spatially localized at the lower-edge vortex, whereas the signal at the upper-edge vortex vanishes within numerical precision (see \cref{Fig2}(c)).
This spatial selectivity follows from the bias-driven quasiparticle flow in~\cref{Fig2}(b): the lower-edge vortex is reached by a nonequilibrium branch originating from the biased left contact, whereas the branch incident on the upper-edge vortex is equilibrated with the grounded TSC and right contact.}
To verify the physical picture illustrated in~\cref{Fig1}(b), we introduce a spatially resolved nonequilibrium occupation function $F_{\mathrm {neq}}^{\mathrm{e}/\mathrm{h}}(\bm{r},E)$. 
Owing to the particle-hole symmetry of the BdG Hamiltonian in~\cref{eq:BdG}, it is sufficient to consider only the electron component, which is defined as
\begin{equation}
F_{\mathrm{neq}}^{\mathrm{e}}(\bm{r}, E) = - i \, G^{<}_{\mathrm{ee}}(\bm{r}, \bm{r}, E) / A_{\mathrm{ee}}(\bm{r}, \bm{r}, E).
\end{equation}

We first examine the local occupation function of the incident QAH edge states (marked by the red star in~\cref{Fig2}(c)), which are expected to remain in thermal equilibrium with the source contact under reflectionless contacts.
This is confirmed by the numerical result in~\cref{Fig2}(e), which exhibits a perfect Fermi distribution at temperature $k_{\mathrm{B}} T\! =\! 0$.
Next, we analyze the evolution of the local occupation function, $F_{\mathrm{neq}}^{\mathrm{e}}(\bm{r}, E)$, along the line cut traversing the vortex core (marked by the dashed red line in~\cref{Fig2}(c)). As shown in~\cref{Fig2}(f), the numerical result reveals a transition from nonequilibrium to equilibrium occupation as the CMEM propagates across the vortex.
This irreversible relaxation process is accompanied by energy dissipation, i.e., non-zero $\dot{Q}$ in~\cref{Fig2}(c).
As an additional control, we consider a configuration in which a vortex is placed within the bulk region of the TSC, far from the edge states. As a result, no energy dissipation is observed in~\cref{Fig2}(d).
{These results show that coupling to a vortex core alone is not sufficient for energy dissipation. The vortex must also overlap with a nonequilibrium quasiparticle flow.}

\section{Dissipation in realistic samples}

\begin{figure}[!htbp]
  \centering
  \includegraphics[width=0.88\linewidth]{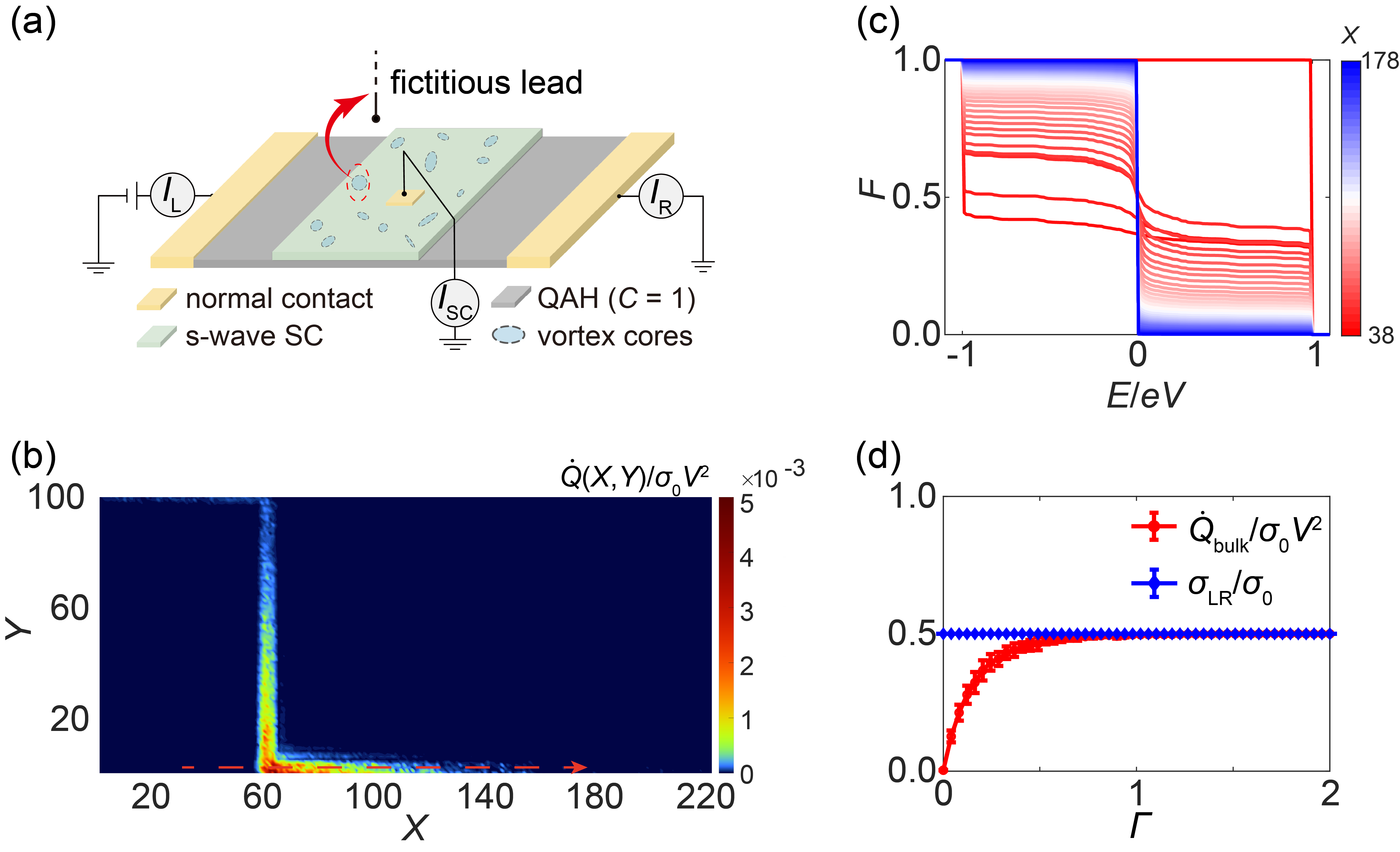}
  \caption{
    {Spatial dissipation pattern in a realistic chiral TSC with $N\!=\!1$ in a voltage-biased electrical-transport setup.}
    (a) Schematic of a device containing vortices with normal cores randomly distributed in the device.
    (b) Spatial pattern of the local energy dissipation power $\dot{Q}(X,Y)$, expressed in units of $\sigma_0 V^2$ (with $\sigma_0 = e^2/h$), averaged over 560 random vortex configurations.  
    (c) Local occupation function $F$ along the dashed red line in panel (b).
    (d) Total dissipation power $\dot{Q}_{\mathrm{bulk}}$ (red) and two-terminal conductance $\sigma_{\mathrm{LR}}$ (blue) as functions of the relaxation strength $\Gamma$.
  }
  \label{Fig3}
\end{figure}

{In realistic samples, static pinned Abrikosov vortices or superconductivity-suppressed regions with normal metallic cores may be distributed throughout the device (\cref{Fig3}(a)).}
We model the vortices as fictitious leads characterized by a self-energy term $\Sigma_{\varphi}$. For phenomenological purposes, we model the retarded self-energy as 
\(
\Sigma_{\varphi}(\bm{r}_p)\!=\! -i {\Gamma}/{2} \delta(\bm{r} - \bm{r}_p),
\)
where $\Gamma \!\propto\! 1/\tau$ represents the effective dissipation strength, with $\tau$ being the relaxation time associated with the fictitious lead~\cite{sadasivamThermalTransportMetal2017}. In the context of electron-phonon interactions in metals, the relaxation time is typically temperature-dependent, with $1/\tau_{\mathrm{el-ph}} \!\sim\! T^3$ in the low-temperature limit~\cite{ashcroftSolidStatePhysics2021}.

Energy dissipation is observed along both CMEMs that emerge at the \( N\!=\!1\) TSC boundary in~\cref{Fig3}(b).
It agrees with the quasiparticle current trajectories in~\cref{Fig2}(b). This behavior is further manifested in the local occupation function plotted in~\cref{Fig3}(c), which captures the nonequilibrium evolution along the bottom edge (marked by the dashed red line in~\cref{Fig3}(b)).
In the QAH region, the incident current retains a Fermi distribution as expected ($X \!\sim\! 38$ in~\cref{Fig3}(c)), and thus no energy dissipation occurs even with the presence of fictitious leads.
In contrast, within the TSC region, scattering processes between the CMEMs and vortices induce a nonequilibrium occupation.
Due to inelastic scattering modeled by $\Sigma_{\varphi}$, the nonequilibrium state gradually relaxes to an equilibrium distribution sharing the same chemical potential as the superconductor (\cref{Fig3}(c) for $X \!>\! 60$).
During this relaxation process, energy dissipation emerges along the path of edge-state transport.

\begin{figure}[!htbp]
  \centering
  \includegraphics[width=0.96\linewidth]{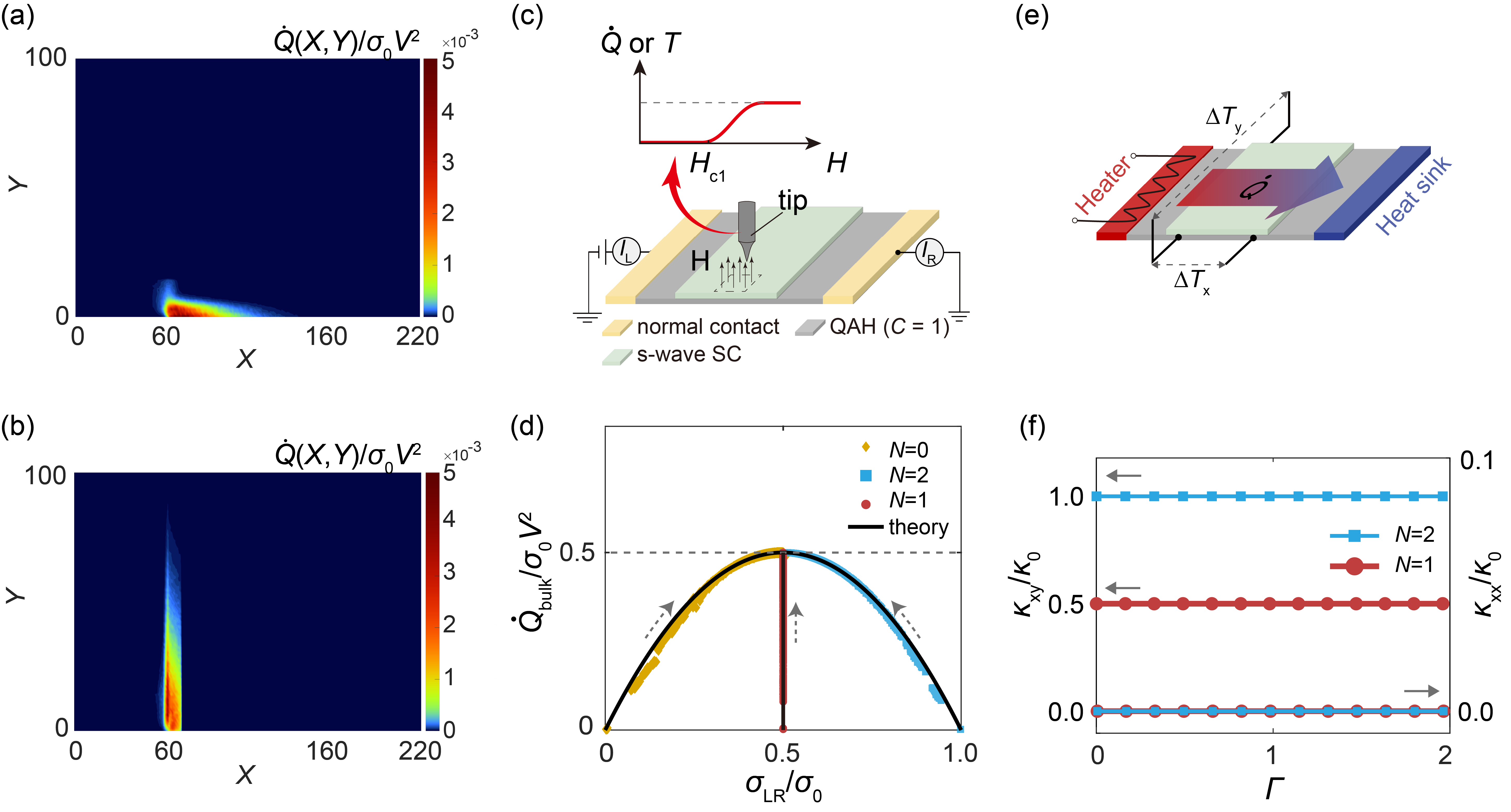}
  \caption{
(a)(b) Spatial maps of the local energy dissipation power $\dot{Q}(X,Y)$ for QAH--TSC--QAH junctions with central regions at Chern numbers $N\!=\!2$ and $N\!=\!0$, respectively.  
(c) Optional experimental control for the nanothermometry measurement.
In a type-II superconducting film, a perpendicular field $H$ tunes the mean density of pinned Abrikosov vortices in the mixed state.
(d) Parametric plot of the total dissipation power $\dot{Q}_{\mathrm{bulk}}$ versus the two-terminal conductance $\sigma_{\mathrm{LR}}$, constructed from the $\Gamma$-dependent numerical data. The solid line denotes the semiclassical prediction and the colored points denote the numerical results. The dashed gray arrows indicate increasing $\Gamma$ toward the point $(0.5,0.5)$.
{(e) Schematic of the thermal Hall measurement, driven by a temperature bias under open-circuit electrical conditions.}
(f) Numerical results for the intrinsic electronic thermal Hall conductance $\kappa_{xy}$ {(left axis)} and the longitudinal thermal conductance $\kappa_{xx}$ {(right axis)}, both normalized by the quantum of thermal conductance $\kappa_0\!=\!\pi^2 k_{\mathrm{B}}^2 T_0/3h$, as functions of the relaxation strength $\Gamma$ in the $N\!=\!1$ and $N\!=\!2$ TSC phases.
}
  \label{Fig4}
\end{figure}

We investigate how the relaxation strength $\Gamma$ influences the two-terminal conductance $\sigma_{\mathrm{LR}}$ and the dissipation power $\dot{Q}_{\mathrm{bulk}}$ within the device. {Following the conductance convention of Wang et al.~\cite{wangChiralTopologicalSuperconductor2015}, we define the source-drain conductance as $\sigma_{\mathrm{LR}}\equiv(I_L-I_R)/(2V)$ for $V_L=V$ and $V_R=V_S=0$. } The total dissipation power is $\dot{Q}_{\mathrm{bulk}}\!=\!\sum_{X,Y}\dot{Q}(X,Y)$.
As shown in~\cref{Fig3}(d), $\sigma_{\mathrm{LR}}$ remains half-quantized across a range of relaxation strengths $\Gamma$, whereas $\dot{Q}_{\mathrm{bulk}}$ increases from 0 and eventually saturates at 0.5 (in units of $\sigma_0 V^2$).
This robust conductance highlights the topological protection of the edge mode in the $N\!=\!1$ phase. However, half-quantized conductance, once widely regarded as a key signature of chiral Majorana fermions~\cite{wangChiralTopologicalSuperconductor2015,chungConductanceNoiseSignatures2011a}, {can also arise from non-Majorana mechanisms~\cite{Chenchuizhen2017PRBdomainwall,JiWenjieWenxiaogang2018PRL,HuangYingyi2018PRB,LianBiao2018PRB,kayyalhaAbsenceEvidenceChiral2020,udayNonMajoranaOriginHalfinteger2025}}.

{Energy dissipation also occurs in the \(N=2\) and \(N=0\) junctions, but with distinct spatial profiles determined by their edge-mode connectivity, as summarized in Supplementary Fig.~1. The corresponding spatial maps are shown in \cref{Fig4}(a,b), respectively. Full counterparts of \cref{Fig3} for \(N=0\) and \(N=2\) are provided in Supplementary Figs.~3 and 4, respectively.} In the \(N=2\) case, the edge current is fully transmitted through the TSC region, and dissipation develops along the bottom edge (\cref{Fig4}(a)), consistent with the topological equivalence between a \(C=1\) QAH state and an \(N=2\) TSC.
In the \(N=0\) case, the edge current is fully reflected at the QAH--TSC interface, and dissipation follows the vertical interface (\cref{Fig4}(b)). Unlike the \(N=1\) case, the two-terminal conductance in the \(N=0\) (Supplementary Fig.~3) and \(N=2\) (Supplementary Fig.~4) junctions evolves toward \(0.5\sigma_0\) as the relaxation strength \(\Gamma\) increases.

To compare energy dissipation with electrical transport, \cref{Fig4}(d) replots the $\Gamma$-dependent numerical data parametrically as $\dot{Q}_{\mathrm{bulk}}$ versus $\sigma_{\mathrm{LR}}$, with $\Gamma$ increasing along the dashed gray arrows. 
Through a semiclassical analysis (see Supplementary Note~6), we derive a relation between the dimensionless dissipation power $\tilde{\dot{Q}}_{\mathrm{bulk}}$ ($\equiv \dot{Q}_{\mathrm{bulk}} / \sigma_0 V^2$) and the dimensionless two-terminal conductance $\tilde{\sigma}_{\mathrm{LR}}$ ($\equiv \sigma_{\mathrm{LR}} / \sigma_0$) for $N\!=\!0$ and $N\!=\!2$ cases, given by:
\begin{equation}
\tilde{\dot{Q}}_{\mathrm{bulk}}  =  2 \tilde{\sigma}_{\mathrm{LR}} \left( 1 - \tilde{\sigma}_{\mathrm{LR}} \right),
\label{eq:QandS}
\end{equation}
where $\sigma_0 = e^2 / h$ is the conductance quantum, and $V$ is the voltage across the junction.
For the $N\!=\!1$ case, where $\sigma_{\mathrm{LR}}$ remains fixed at $0.5\sigma_0$, Eq.~\eqref{eq:QandS} provides the upper bound for the dissipation power.
The numerical results for the $N\!=\!0$ and $N\!=\!2$ junctions closely follow the analytical prediction in~\cref{Fig4}(d), thereby validating the semiclassical relation.

A joint analysis of charge transport and energy dissipation yields two observations: (i) The value $\sigma_\mathrm{LR}\!=\!0.5\sigma_0$ is shared across the three TSC phases, indicating that it is not a unique electrical fingerprint of any particular phase;
(ii) $N\!=\!2$ and $N\!=\!0$ phases can be distinguished from the $N\!=\!1$ phase of particular interest by the distinct spatial dissipation profiles in Figs.~\ref{Fig3}(b),~\ref{Fig4}(a) and~\ref{Fig4}(b), serving as a fingerprint to resolve this ambiguity.

\section{Discussion}
Our findings {offer a way to address the ambiguity of} the half-quantized conductance plateau. While originally predicted as a hallmark of chiral Majorana modes~\cite{wangChiralTopologicalSuperconductor2015}, recent experiments (e.g., Kayyalha \textit{et al.}~\cite{kayyalhaAbsenceEvidenceChiral2020}) demonstrated that a topologically distinct $C=1$ QAH phase (equivalent to an $N=2$ chiral TSC) mimics this signature. They attributed this observation to the superconducting condensate acting as an electrical short, which masks the specific bulk topology. Our theoretical results for the $N=2$ junction (Fig.~\ref{Fig4}(a)) quantitatively reproduce this scenario. We confirm that in this effective $N=2$ configuration, the superconducting condensate indeed shorts the chiral edge transport, leading to the same half-quantized conductance observed in experiments.

Crucially, however, we show that dissipation imaging breaks this degeneracy. Although the experimentally realized $N=2$ phase and the intrinsic $N=1$ chiral TSC share identical electrical conductance, their spatial dissipation profiles differ fundamentally. The $N=1$ phase features dissipation tracing both split Majorana branches (Fig.~\ref{Fig3}(b)), whereas the $N=2$ phase dissipates exclusively at the downstream boundary (Fig.~\ref{Fig4}(a)). This contrast shows that spatial dissipation distinguishes the \(N\!=\!1\) and \(N\!=\!2\) edge-connectivity patterns even when their electrical conductances are identical.

{To test whether a trivial \(N\!=\!0\) device can reproduce the \(N\!=\!1\) spatial signature, we considered nontopological edge channels along the device boundary (Supplementary Fig.~2). Generic edge disorder localizes the trivial edge channels and suppresses the dissipation along the bottom edge. Thus, these trivial edge channels cannot mimic the connected \(N\!=\!1\) dissipation pattern, even when the conductance and total dissipated power remain half-quantized.}

{As a complementary consistency check, we examine whether the vortex-induced equilibration included in our model is compatible with the conventional chiral-edge thermal Hall response. Unlike the voltage-driven dissipation calculations, the setup in Fig.~\ref{Fig4}(e) is driven by a temperature difference under open-circuit electrical conditions, while all thermometric contacts and fictitious vortex probes are floating and adiabatic. These probes may equilibrate the local quasiparticle distribution but cannot absorb or release net heat. 
Local inelastic equilibration, by itself, does not alter the chirality of the edge channels or generate counterpropagating heat-transport paths.
Thus, the \(N=1,~2\) phases have a vanishing longitudinal thermal response, \(\kappa_{xx}=0\), and a quantized thermal Hall conductance, \(\kappa_{xy}=N\kappa_0/2\) (see Supplementary Note~7 for an analytical proof). The numerical results in Fig.~\ref{Fig4}(f) therefore give \(\kappa_{xy}=0.5\kappa_0\) and \(\kappa_0\) for the \(N=1\) and \(N=2\) phases, respectively (see Supplementary Note~5 for numerical details). Finite nonuniversal longitudinal or Hall backgrounds in realistic devices, arising from bulk quasiparticles, phonons, additional edge channels, or substrate heat leakage, are outside the intrinsic electronic response considered here.}

\begin{figure}[!htbp]
    \centering
    \includegraphics[width=0.82\linewidth]{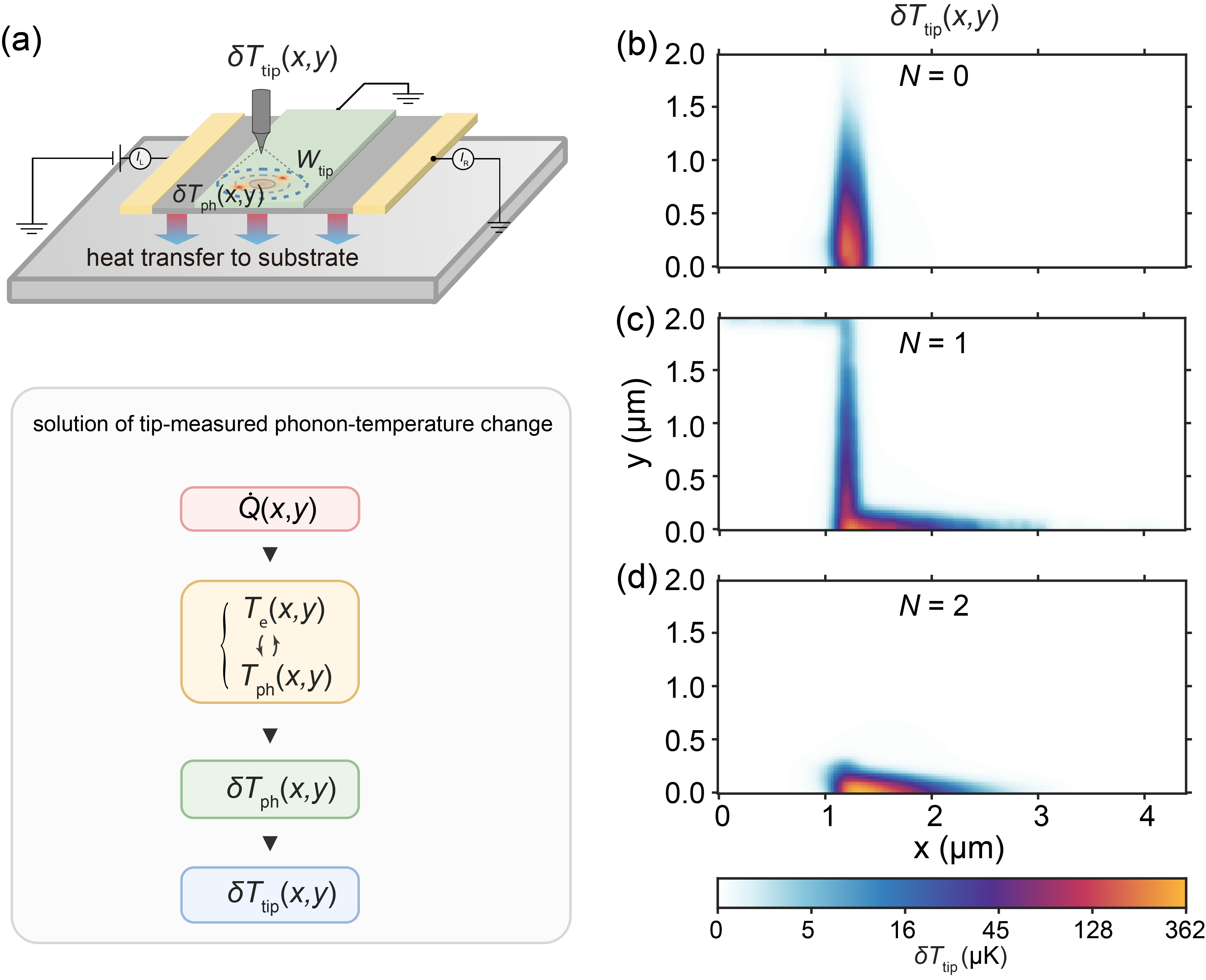}
    \caption{
    (a) Schematic of scanning superconducting quantum interference device
    (SQUID)-on-tip thermometry and numerical procedure used to convert the NEGF dissipation
    density \(\dot{Q}(x,y)\) into the tip-measured temperature rise
    \(\delta T_{\mathrm{tip}}(x,y)\). The stationary electron--phonon
    two-temperature model includes in-plane heat conduction,
    electron--phonon energy exchange, film--substrate heat transfer,
    three-dimensional spreading through a finite-thickness substrate,
    and convolution with a Gaussian point-spread function of
    \(50\,\mathrm{nm}\) full width at half maximum.
    (b)--(d) Calculated \(\delta T_{\mathrm{tip}}(x,y)\) for
    (b) \(N=0\), (c) \(N=1\), and (d) \(N=2\), at
    \(V=0.1\,\mathrm{mV}\) and \(T_0=50\,\mathrm{mK}\).
    The \(N=1\) map retains a vertical-interface branch that is absent
    in the predominantly lower-edge \(N=2\) map. 
    }
    \label{fig:dissipation_to_temperature}
\end{figure}

{To assess experimental feasibility after heat spreading, we convert the voltage-biased NEGF dissipation maps into the phonon-temperature rise detectable for a finite-size SQUID-on-tip thermometer, using a steady-state two-temperature model for superconducting films and nanowires~\cite{semenovAnalysisNonequilibriumPhotoresponse1995a,daneSelfheatingHotspotsSuperconducting2022}.
The calculation includes electronic and phononic in-plane heat diffusion, electron--phonon energy exchange, heat transfer through the substrate, and finite-tip spatial averaging (Supplementary Note~5). We use a bias \(V=0.1\,\mathrm{mV}\) and a base temperature
\(T_0=50\,\mathrm{mK}\). For
\(\sigma_{\mathrm{LR}}=0.5e^2/h\), it corresponds to a current \(I\simeq1.9\,\mathrm{nA}\), comparable to the excitation currents in recent QAH--superconductor transport experiments~\cite{kayyalhaAbsenceEvidenceChiral2020,
udayNonMajoranaOriginHalfinteger2025}.
For experimentally motivated parameters, the tip-measured temperature rise reaches approximately \(3.6\times10^2\,\mu\mathrm{K}\), with extended signals in the tens-to-hundreds-of-microkelvin range (Fig.~\ref{fig:dissipation_to_temperature}). Substrate-mediated heat spreading and finite-tip averaging broaden the temperature profiles but preserve their distinct spatial pattern. The \(N\!=\!0\), \(N\!=\!1\), and \(N\!=\!2\) phases exhibit interface-localized (\cref{fig:dissipation_to_temperature}(b)), L-shaped (\cref{fig:dissipation_to_temperature}(c)), and predominantly lower-edge temperature patterns (\cref{fig:dissipation_to_temperature}(d)), respectively.}

{These results motivate a direct experimental validation using scanning
SQUID-on-tip nanothermometry. This technique has demonstrated
sub-\(50\,\mathrm{nm}\) spatial sensitivity and sub-microkelvin temperature sensitivity
~\cite{halbertalNanoscaleThermalImaging2016}, while its operation with
thermal-imaging capability in a cryogen-free dilution refrigerator has
also been established
~\cite{zhouScanningSQUIDontipMicroscope2023}.}
As an additional experimental control, a perpendicular field may be used to tune the mean density of pinned Abrikosov vortices in the mixed-state regime of a type-II superconducting film, such as thin Nb, as illustrated in Fig.~\ref{Fig4}(c).
In this regime, the mean magnetic induction $\bar B$ sets $n_v\simeq \bar B/\Phi_0$, with $\Phi_0=h/2e$, whereas pinning determines the microscopic vortex positions.
This field dependence provides an experimental control over the dissipation strength, providing an independent consistency check on the vortex-induced origin of the measured signal.

\section{Conclusion}
We have established that spatially resolved energy dissipation serves as a fingerprint for chiral topological superconductivity, {helping to resolve} the long-standing ambiguity of electrical conductance measurements. By demonstrating that static vortices induce topologically distinct dissipation profiles, our work provides {a criterion to help distinguish} chiral Majorana modes from trivial mimics. Backed by the topological robustness of the thermal Hall conductance, these findings establish dissipation, rather than merely as an undesirable effect, as a powerful diagnostic tool.
Our proposal provides {a concrete experimental route for probing} chiral Majorana fermions using existing nanothermometry techniques.

\section*{Acknowledgements}
We thank Qing-Feng Sun, Haiwen Liu, and Yijia Wu for illuminating discussions.
This work was financially supported by the National Key R\&D Programme of China (grant No. 2024YFA1409003 and No. 2022YFA1403700), National Natural Science Foundation of China (No. 12634001, No. 12674053, No. 12350401 and No. 12447123), the Shanghai Science and Technology Innovation Action Plan (No. 24LZ1400800), Quantum Science and Technology-National Science and Technology Major Project (Grants No. 2021ZD0302400), the China Postdoctoral Science Foundation (Grant No. 2024M760491) and Shanghai Oriental Talent Program Youth Project (Grant No. QNJY202508). H. Li was also supported by the Chenguang Program of Shanghai Education Development Foundation and Shanghai Municipal Education Commission.

\section*{Conflict of interest}
The authors declare that they have no conflict of interest.


{\small
\begin{thebibliography}{10}

\bibitem{kitaevUnpairedMajoranaFermions2001}
A.~Y. Kitaev,
\newblock Phys.-Usp. {\bfseries 44}, 131 (2001).

\bibitem{hasanColloquiumTopologicalInsulators2010}
M.~Z. Hasan, and C.~L. Kane,
\newblock Rev. Mod. Phys. {\bfseries 82}, 3045 (2010).

\bibitem{qiTopologicalInsulatorsSuperconductors2011}
X.-L. Qi, and S.-C. Zhang,
\newblock Rev. Mod. Phys. {\bfseries 83}, 1057 (2011).

\bibitem{aliceaNewDirectionsPursuit2012}
J.~Alicea,
\newblock Rep. Prog. Phys. {\bfseries 75}, 076501 (2012).

\bibitem{nadj-perge2014ScienceMajoranaFermionsFeChain}
S.~{Nadj-Perge}, I.~K. Drozdov, J.~Li, H.~Chen, S.~Jeon, J.~Seo, A.~H. MacDonald, B.~A. Bernevig, and A.~Yazdani,
\newblock Science {\bfseries 346}, 602 (2014).

\bibitem{mourikSignaturesMajoranaFermions2012}
V.~Mourik, K.~Zuo, S.~M. Frolov, S.~R. Plissard, E.~P. A.~M. Bakkers, and L.~P. Kouwenhoven,
\newblock Science {\bfseries 336}, 1003 (2012).

\bibitem{sunJiaJF2016PRLMajoranaZeroMode}
H.-H. Sun, K.-W. Zhang, L.-H. Hu, C.~Li, G.-Y. Wang, H.-Y. Ma, Z.-A. Xu, C.-L. Gao, D.-D. Guan, Y.-Y. Li, C.~Liu, D.~Qian, Y.~Zhou, L.~Fu, S.-C. Li, F.-C. Zhang, and J.-F. Jia,
\newblock Phys. Rev. Lett. {\bfseries 116}, 257003 (2016).

\bibitem{zhangDingHong2018ScienceIronFeTopologicalSuperconductivity}
P.~Zhang, K.~Yaji, T.~Hashimoto, Y.~Ota, T.~Kondo, K.~Okazaki, Z.~J. Wang, J.~S. Wen, G.~D. Gu, H.~Ding, and S.~Shin,
\newblock Science {\bfseries 360}, 182 (2018).

\bibitem{wangDingH2018ScienceEvidenceMajoranaIronTopologicalSuperconductivity}
D.~F. Wang, L.~Y. Kong, P.~Fan, H.~Chen, S.~Y. Zhu, W.~Y. Liu, L.~Cao, Y.~J. Sun, S.~X. Du, J.~Schneeloch, R.~D. Zhong, G.~D. Gu, L.~Fu, H.~Ding, and H.~J. Gao,
\newblock Science {\bfseries 362}, 333 (2018).

\bibitem{FuliangKane2009PRL}
L.~Fu, and C.~L. Kane,
\newblock Phys. Rev. Lett. {\bfseries 102}, 216403 (2009).

\bibitem{AkhmerovBeenakker2009PRL}
A.~R. Akhmerov, J.~Nilsson, and C.~W.~J. Beenakker,
\newblock Phys. Rev. Lett. {\bfseries 102}, 216404 (2009).

\bibitem{Beenakker2013annurev-conmatphys}
C.~W.~J. Beenakker,
\newblock Annu. Rev. Condens. Matter Phys. {\bfseries 4}, 113 (2013).

\bibitem{Elliott2015RevModPhys}
S.~R. Elliott, and M.~Franz,
\newblock Rev. Mod. Phys. {\bfseries 87}, 137 (2015).

\bibitem{qiChiralTopologicalSuperconductor2010}
X.-L. Qi, T.~L. Hughes, and S.-C. Zhang,
\newblock Phys. Rev. B {\bfseries 82}, 184516 (2010).

\bibitem{yanqing2021CommunPhysAxionMajorana}
Q.~Yan, H.~Li, J.~Zeng, Q.-F. Sun, and X.~C. Xie,
\newblock Commun. Phys. {\bfseries 4}, 239 (2021).

\bibitem{lianTopologicalQuantumComputation2018a}
B.~Lian, X.-Q. Sun, A.~Vaezi, X.-L. Qi, and S.-C. Zhang,
\newblock Proc. Natl. Acad. Sci. U.S.A. {\bfseries 115}, 10938 (2018).

\bibitem{zhouNonAbelianOperationChiral2019}
Y.-F. Zhou, Z.~Hou, and Q.-F. Sun,
\newblock Phys. Rev. B {\bfseries 99}, 195137 (2019).

\bibitem{linNonAbelianOperationScattering2022}
Z.-X. Lin, Y.~Wu, and X.~C. Xie,
\newblock Phys. Rev. B {\bfseries 105}, 205428 (2022).

\bibitem{wuRecentProgressNonAbelian2023}
Y.~Wu, J.~Liu, and X.~C. Xie,
\newblock Sci. China-Phys. Mech. Astron. {\bfseries 66}, 267004 (2023).

\bibitem{huangUnifiedModelNonAbelian2026}
T.~Huang, R.~Zhang, X.~Li, X.-J. Liu, X.~C. Xie, and Y.~Wu,
\newblock Sci. China-Phys. Mech. Astron. {\bfseries 69}, 227211 (2026).

\bibitem{Nayak2008RevModPhys}
C.~Nayak, S.~H. Simon, A.~Stern, M.~Freedman, and S.~Das~Sarma,
\newblock Rev. Mod. Phys. {\bfseries 80}, 1083 (2008).

\bibitem{Campbell2017naturereviewQC}
E.~T. Campbell, B.~M. Terhal, and C.~Vuillot,
\newblock Nature {\bfseries 549}, 172 (2017).

\bibitem{chungConductanceNoiseSignatures2011a}
S.~B. Chung, X.-L. Qi, J.~Maciejko, and S.-C. Zhang,
\newblock Phys. Rev. B {\bfseries 83}, 100512(R) (2011).

\bibitem{wangChiralTopologicalSuperconductor2015}
J.~Wang, Q.~Zhou, B.~Lian, and S.-C. Zhang,
\newblock Phys. Rev. B {\bfseries 92}, 064520 (2015).

\bibitem{Chenchuizhen2017PRBdomainwall}
C.-Z. Chen, J.~J. He, D.-H. Xu, and K.~T. Law,
\newblock Phys. Rev. B {\bfseries 96}, 041118(R) (2017).

\bibitem{JiWenjieWenxiaogang2018PRL}
W.~Ji, and X.-G. Wen,
\newblock Phys. Rev. Lett. {\bfseries 120}, 107002 (2018).

\bibitem{HuangYingyi2018PRB}
Y.~Huang, F.~Setiawan, and J.~D. Sau,
\newblock Phys. Rev. B {\bfseries 97}, 100501(R) (2018).

\bibitem{kayyalhaAbsenceEvidenceChiral2020}
M.~Kayyalha, D.~Xiao, R.~Zhang, J.~Shin, J.~Jiang, F.~Wang, Y.-F. Zhao, R.~Xiao, L.~Zhang, K.~M. Fijalkowski, P.~Mandal, M.~Winnerlein, C.~Gould, Q.~Li, L.~W. Molenkamp, M.~H.~W. Chan, N.~Samarth, and C.-Z. Chang,
\newblock Science {\bfseries 367}, 64 (2020).

\bibitem{udayNonMajoranaOriginHalfinteger2025}
A.~Uday, G.~Lippertz, B.~Bhujel, A.~A. Taskin, and Y.~Ando,
\newblock Phys. Rev. B {\bfseries 111}, 035440 (2025).

\bibitem{royHalfintegerThermalConductance2026}
U.~Roy, S.~Manna, S.~Chakraborty, K.~Watanabe, T.~Taniguchi, A.~Das, M.~Goldstein, Y.~Gefen, and A.~Das,
\newblock Nat. Commun. {\bfseries 17}, 2853 (2026).

\bibitem{changColloquiumQuantumAnomalous2023}
C.-Z. Chang, C.-X. Liu, and A.~H. MacDonald,
\newblock Rev. Mod. Phys. {\bfseries 95}, 011002 (2023).

\bibitem{WangjingZhangSC2017NatMaterTopoCondensedMatter}
J.~Wang, and S.-C. Zhang,
\newblock Nat. Mater. {\bfseries 16}, 1062 (2017).

\bibitem{slizovskiyCoolingChiralHeat2017}
S.~Slizovskiy, and V.~Fal'ko,
\newblock Phys. Rev. B {\bfseries 96}, 075434 (2017).

\bibitem{margueriteImagingWorkDissipation2019}
A.~Marguerite, J.~Birkbeck, A.~{Aharon-Steinberg}, D.~Halbertal, K.~Bagani, I.~Marcus, Y.~Myasoedov, A.~K. Geim, D.~J. Perello, and E.~Zeldov,
\newblock Nature {\bfseries 575}, 628 (2019).

\bibitem{zhangDissipationResistanceImaging2020}
G.~Zhang, I.~V. Gornyi, and A.~D. Mirlin,
\newblock Phys. Rev. Res. {\bfseries 2}, 013337 (2020).

\bibitem{fangThermalDissipationQuantum2021}
J.-Y. Fang, N.-X. Yang, Q.~Yan, A.-M. Guo, and Q.-F. Sun,
\newblock Phys. Rev. B {\bfseries 104}, 115411 (2021).

\bibitem{liEmergentEnergyDissipation2024}
H.~Li, H.~Jiang, Q.-F. Sun, and X.~C. Xie,
\newblock Sci. Bull. {\bfseries 69}, 1221 (2024).

\bibitem{yanRulesDissipationlessTopotronics2024}
Q.~Yan, H.~Li, H.~Jiang, Q.-F. Sun, and X.~C. Xie,
\newblock Sci. Adv. {\bfseries 10}, eado4756 (2024).

\bibitem{yanDissipationlessTopotronics2025}
Q.~Yan, H.~Li, H.~Jiang, and X.~C. Xie,
\newblock Mater. Today Quantum {\bfseries 5}, 100023 (2025).

\bibitem{liNewAdvancesDissipationless2025}
H.~Li, Q.~Yan, H.~Jiang, and X.~C. Xie,
\newblock Chin. Sci. Bull. {\bfseries 70}, 1 (2025).

\bibitem{zhouNonHermitianTopology2026}
H.~Zhou, M.~Lu, C.-Z. Chen, and X.~C. Xie,
\newblock Nat. Commun. {\bfseries 17}, 10259 (2026).

\bibitem{weiHiddenDissipation2026}
M.~Wei, B.~Wang, F.~Xu, and J.~Wang,
\newblock Phys. Rev. Lett. {\bfseries 137}, 046303 (2026).

\bibitem{liuDissipationDrivenFirstOrder2026}
X.~Liu, M.~Lu, H.~Liu, and X.~C. Xie,
\newblock Sci. China-Phys. Mech. Astron. {\bfseries 69}, 267411 (2026).

\bibitem{bardeenTheoryMotionVortices1965}
J.~Bardeen, and M.~J. Stephen,
\newblock Phys. Rev. {\bfseries 140}, A1197 (1965).

\bibitem{Altland-ZirnbauerSymmetryClassificationPRB}
A.~P. Schnyder, S.~Ryu, A.~Furusaki, and A.~W.~W. Ludwig,
\newblock Phys. Rev. B {\bfseries 78}, 195125 (2008).

\bibitem{halbertalNanoscaleThermalImaging2016}
D.~Halbertal, J.~Cuppens, M.~B. Shalom, L.~Embon, N.~Shadmi, Y.~Anahory, H.~R. Naren, J.~Sarkar, A.~Uri, Y.~Ronen, Y.~Myasoedov, L.~S. Levitov, E.~Joselevich, A.~K. Geim, and E.~Zeldov,
\newblock Nature {\bfseries 539}, 407 (2016).

\bibitem{zhouScanningSQUIDontipMicroscope2023}
H.~Zhou, N.~Auerbach, I.~Roy, M.~Bocarsly, M.~E. Huber, B.~Barick, A.~Pariari, M.~H{\"u}cker, Z.~S. Lim, A.~Ariando, A.~I. Berdyugin, N.~Xin, M.~Rappaport, Y.~Myasoedov, and E.~Zeldov,
\newblock Rev. Sci. Instrum. {\bfseries 94}, 053706 (2023).

\bibitem{yanqing2019PRBCMEMs}
Q.~Yan, Y.-F. Zhou, and Q.-F. Sun,
\newblock Phys. Rev. B {\bfseries 100}, 235407 (2019).

\bibitem{QiWuZhang2006PRBQAHmodel}
X.-L. Qi, Y.-S. Wu, and S.-C. Zhang,
\newblock Phys. Rev. B {\bfseries 74}, 085308 (2006).

\bibitem{savichQuasiparticleRelaxation2017}
Y.~Savich, L.~Glazman, and A.~Kamenev,
\newblock Phys. Rev. B {\bfseries 96}, 104510 (2017).

\bibitem{samokhvalovElectronPhonon2022}
A.~V. Samokhvalov, I.~A. Shereshevskii, N.~K. Vdovicheva, M.~Taupin, I.~M. Khaymovich, and A.~S. Mel'nikov,
\newblock Phys. Rev. B {\bfseries 105}, 024522 (2022).

\bibitem{vaitkusButtikerProbesRecursive2018}
J.~A. Vaitkus, and J.~H. Cole,
\newblock Phys. Rev. B {\bfseries 97}, 085149 (2018).

\bibitem{maianiConductanceMatrixSymmetries2022}
A.~Maiani, M.~Geier, and K.~Flensberg,
\newblock Phys. Rev. B {\bfseries 106}, 104516 (2022).

\bibitem{mortensenDephasingSemiconductorSuperconductor2000}
N.~A. Mortensen, A.-P. Jauho, and K.~Flensberg,
\newblock Superlattices Microstruct. {\bfseries 28}, 67 (2000).

\bibitem{mishraMajoranaFermionInduced2024}
S.~Mishra, R.~Das, and C.~Benjamin,
\newblock J. Appl. Phys. {\bfseries 136}, 224304 (2024).

\bibitem{sadasivamThermalTransportMetal2017}
S.~Sadasivam, N.~Ye, J.~P. Feser, J.~Charles, K.~Miao, T.~Kubis, and T.~S. Fisher,
\newblock Phys. Rev. B {\bfseries 95}, 085310 (2017).

\bibitem{ashcroftSolidStatePhysics2021}
N.~W. Ashcroft, and N.~D. Mermin,
\newblock {\itshape Solid State Physics} (Cengage, Boston, MA, 2021).

\bibitem{LianBiao2018PRB}
B.~Lian, J.~Wang, X.-Q. Sun, A.~Vaezi, and S.-C. Zhang,
\newblock Phys. Rev. B {\bfseries 97}, 125408 (2018).

\bibitem{semenovAnalysisNonequilibriumPhotoresponse1995a}
A.~D. Semenov,
\newblock Phys. Rev. B {\bfseries 52}, 581 (1995).

\bibitem{daneSelfheatingHotspotsSuperconducting2022}
A.~Dane, J.~Allmaras, D.~Zhu, M.~Onen, M.~Colangelo, R.~Baghdadi, J.-L. Tambasco, Y.~Morimoto, I.~E. Forno, I.~Charaev, Q.~Zhao, M.~Skvortsov, A.~Kozorezov, and K.~K. Berggren,
\newblock Nat. Commun. {\bfseries 13}, 5429 (2022).

\end{thebibliography}

}
\end{document}